\documentclass[12pt,eadjoint tfnpsf]{article}

\catcode`\@=11
\@addtoreset{equation}{section}

\global\arraycolsep=1pt

\usepackage{amsbsy,amssymb,latexsym,amsfonts, amsmath}
\usepackage{mathrsfs}
\usepackage{graphicx}

\RequirePackage[dvips,usenames]{color}
\definecolor{fireblick}{rgb}{0.698039,0.133333,0.133333}

\newcommand{\beq}{\begin{equation}}
\newcommand{\eeq}{\end{equation}}
\newcommand{\bea}{\begin{eqnarray}}
\newcommand{\eea}{\end{eqnarray}}

\newcommand{\CF}{{\mathcal F}}

\newcommand{\CN}{{\mathcal N}}
\newcommand{\CO}{{\mathcal O}}

\newcommand\tr{\mathrm{tr}}

\renewcommand{\thefootnote}{\fnsymbol{footnote}}
\begin{document}
%%%%%%%%%%%%%%%%%%%%%%%%%%%%%%%%%%%%%%%%%%%%%%%%%%%%%%%%%%%%%%%%%%%%%%%%%%%%%%%%%%%%%%%%%%
\begin{titlepage}
%%%%%%%%%%%%%%%%%%%% preprint # %%%%%%%%%%%%%%%%%%

\begin{flushright}
\normalsize
%\filename
~~~~
YITP-10-40\\
\today\\
\end{flushright}
%%%%%%%%%%%%%%%%%%%%%%%%%%%%%%%%%%%%%%%%%%%%%%%%%%

\vspace{80pt}

%%%%%%%%%%%%%%%%%%%% title %%%%%%%%%%%%%%%%%%%%%%%
\begin{center}
{\Large Seiberg-Witten theory, matrix model and AGT relation}\\
\end{center}
%%%%%%%%%%%%%%%%%%%%%%%%%%%%%%%%%%%%%%%%%%%%%%%%%%

\vspace{25pt}

%%%%%%%%%%%%%%%%%%% authors %%%%%%%%%%%%%%%%%%%%%%
\begin{center}
{
Tohru Eguchi
and
Kazunobu Maruyoshi 
}\\
%%%%%%%%%%%%%%%%%%%%%%%%%%%%%%%%%%%%%%%%%%%%%%%%%%
%
\vspace{15pt}
%
%%%%%%%%%%%%%%%%%%% affiliation %%%%%%%%%%%%%%%%%%%
\it Yukawa Institute for Theoretical Physics, Kyoto University, Kyoto 606-8502, Japan\\
\end{center}
%%%%%%%%%%%%%%%%%%%%%%%%%%%%%%%%%%%%%%%%%%%%%%%%%%%
%
\vspace{20pt}
\begin{center}
Abstract\\
\end{center}

We discuss the Penner-type matrix model which has been proposed to explain the AGT relation 
between the 2-dimensional Liouville theory and 4-dimensional $\CN=2$ superconformal gauge theories. 
In our previous communication we have obtained the spectral curve of the matrix model 
and showed that it agrees with that derived from M-theory. 
We have also discussed the decoupling limit of massive flavors and proposed new matrix models 
which describe Seiberg-Witten theory with flavors $N_f=2,3$. 
In this article we explicitly evaluate the free energy of these matrix models 
and show that they in fact reproduce the amplitudes of Seiberg-Witten theory.

%%%%%%%%%%%%%%%%%%%%%%%%%%%%%%%%%%%%%%%%%%%%%%%%%%%

\vfill

\setcounter{footnote}{0}
\renewcommand{\thefootnote}{\arabic{footnote}}

\end{titlepage}

%%%%%%%%%%%%%%%%%%%%%%%%%%%%%%%%%%%%%%%%%%%%%%%%%%%%%%%%%%%%%%%%%%%%%%%%%%%%%%%%%%%%%%%%%%%%%%%%%%%%
\section{Introduction}
\label{sec:intro}

Recently a very interesting relation between the Nekrasov partition function of $\CN=2$ 
conformal invariant $SU(2)$ gauge theory 
and the conformal block of the Liouville field theory was proposed  \cite{AGT}.
It seems that this is the first example of a precise mathematical relationship between  
quantum field theories defined at different space-time dimensions.
There have been various attempts at checking this AGT relation at lower instanton numbers 
by direct evaluation of Liouville correlation functions \cite{MMS1, MMS2, IO, MMMmatrix, MS}.  There have also been  
attempts at proving the relation by comparing the recursion relation satisfied by the descendants of the 
conformal blocks and Nekrasov's partition function \cite{Poghossian, HJS0, FL, HJS}.

On the other hand, a Penner type matrix model has been proposed to interpolate between the Liouville 
theory and gauge theory \cite{DV} and provide an explanation for the AGT relation. In a previous communication \cite{EM} 
we have studied this matrix model and also proposed models for  asymptotically free 
theories obtained by decoupling some of massive flavors.
We have shown that the spectral curves of these matrix models reproduce those based on  the M-theory construction 
and their free energies satisfy the scaling identities known in the $SU(2)$ Seiberg-Witten theory.
(See also \cite{IMO, SWyllard} for $A_{r}$ quiver matrix model).
  
 In this paper we would like to evaluate the free energies of these matrix models in the large $N$ limit explicitly  and show that they in fact exactly reproduce the amplitudes of $SU(2)$ Seiberg-Witten theory. 
  
  In section 2 we first describe the general properties of matrix models. In section 3 we  
  compute the free energies: we integrate the Seiberg-Witten differential of the matrix model and  
  evaluate the filling fraction in terms of the parameters of the spectral curve. 
  We then invert this relation and derive the free energy. 
  We present the computation for $SU(2)$ gauge theory with two, three  
  and four flavors and show that they all reproduce  the amplitudes of Seiberg-Witten theory.
   In section 4 we discuss decoupling limits of some quiver gauge theories. Section 5 is devoted to conclusion and discussion.
  
  Note: in our convention the free energy of the matrix model $F_m$ is off by a factor 4 from 
  that of gauge theory. Thus we will check the agreement $4F_{m}=F_{gauge}$ throughout this paper.
   
%%%%%%%%%%%%%%%%%%%%%%%%%%%%%%%%%%%%%%%%%%%%%%%%%%%%%%%%%%%%%%%%%%%%%%%%%%%%%%%%%%%%%%%%%%%%%%%%%%%%  section
\section{$SU(2)$ gauge theories and matrix models}
\label{sec:matrix}
  It has been proposed that the Nekrasov partition function for $\CN=2$, $SU(2)$ gauge theory with four flavors   
  (summarized in appendix \ref{sec:Nekrasov}) coincide with the four-point conformal block 
  of Liouville theory \cite{AGT}:
    \bea
    Z_{{\rm inst}}^{SU(2)}
     =     Z_{{\rm CFT}}
     \equiv     
           \left< V_{\tilde{m}_\infty} (\infty) V_{\tilde{m}_1}(1) V_{\tilde{m}_2}(q) V_{\tilde{m}_0}(0) \right>.
    \eea
  Here $V_{\tilde{m}}$ is the vertex operator,
  $Q = b + 1/b$ and the central charge of the Liouville theory is $c= 1 + 6Q^2$.
  
  In order to relate the Liouville theory to matrix model, 
  we consider the Dotsenko-Fateev integral representation of the four-point conformal block 
  in terms of the free field $\phi(z)$ \cite{DF}:
    \bea
    Z_{{\rm DF}}
     =     \left<  \left( \int d \lambda_I :e^{b \phi(\lambda_I)} : \right)^N
           V_{\tilde{m}_\infty} (\infty) V_{\tilde{m}_1}(1) V_{\tilde{m}_2}(q) V_{\tilde{m}_0}(0) \right>,
    \eea
  where the vertex operator $V_{\tilde{m}_i}(z_i)$ is given by $:e^{\tilde{m}_i \phi(z_i)}:$ 
  and we have introduced the $N$-fold integration of screening operators. 
  OPE of the scalar field is given by $\phi(z) \phi(\omega) \sim -2 \log (z - \omega)$.
  Momentum conservation condition relates the external momenta and the number of integrals as 
  $\sum_{i=0}^2 \tilde{m}_i + \tilde{m}_\infty + b N = Q$.
  We redefine the momenta as $\tilde{m}_i = \frac{i m_i}{2 g_s}$ for $i = 0, \infty$ 
  and $\tilde{m}_i = \frac{i m_i}{2 g_s} + \frac{Q}{2}$ for $i = 1, 2$ \cite{AGT}.
  Then the above condition becomes
    \bea
    \sum_{i=0}^2 i m_i + i m_\infty + 2 b g_s N
     =     0.
           \label{massrelation4}
    \eea
  As pointed out in \cite{MMM, KMMMP} and recently in \cite{DV} in the context of the AGT relation, 
  the Dotsenko-Fateev representation may be identified as the $\beta$-deformation of a  one matrix integral  
    \bea
    Z_{{\rm DF}}
     =     q^{\frac{m_0 m_2}{2g_s^2}} (1-q)^{\frac{m_1 m_2}{2g_s^2}}
           \left( \prod_{I=1}^N \int d \lambda_I \right) \prod_{I < J} (\lambda_I - \lambda_J)^{-2b^2} 
           e^{\frac{-ib}{g_s} \sum_I W(\lambda_I)}.
           \label{matrix}
    \eea
  In the case of $b=i$, integrations over $\{\lambda_I,I=1,\cdots,N\}$  becomes an integral over a hermitian matrix  $M$ 
  with eigenvalues $\{\lambda_I\}$
  and the action 
    \bea
    W(M)
     =     \sum_{i=0}^2 m_i \log (M - z_i), \hskip1cm z_0=0,\hskip3mm z_1=1,\hskip3mm z_2=q.
           \label{action4}
    \eea
  We identify the parameters $m_i$ with the mass parameters of the corresponding gauge theory.
  The identification of the parameter $b$ with the Nekrasov's deformation parameters is given by 
    \bea
    \epsilon_1
     =   - i b g_s, ~~~
    \epsilon_2
     =   - \frac{i g_s}{b}.
    \eea
  
  In this paper, we focus on the $b=i$ case, i.e. the self-dual background $\epsilon_1 = - \epsilon_2 = g_s$.
  The momentum conservation condition then reduces to
    \bea
    \sum_{i=0}^2 m_i + m_\infty + 2 g_s N
     =     0.
           \label{massrelation4}
    \eea
  This matrix model is expected to reproduce the results of $SU(2)$ gauge theory with $N_f=4$.
  More precisely, as we will see below, the matrix integral  together with the overall factor
 $(1-q)^{\frac{m_1 m_2}{2g_s^2}}$ in (\ref{matrix})
  corresponds to the $SU(2)$ gauge theory. Note that 
  the factor $(1-q)^{\frac{m_1 m_2}{2g_s^2}}$ is the inverse of the $U(1)$ factor discussed in \cite{AGT}.
  (See appendix \ref{sec:Nekrasov}.)
  
  Another point is that the Coulomb moduli parameter $a$ of the gauge theory is identified as
   the filling fraction $g_s N_i$,
  where $N_i$ is a number of screening operators inserted into  
 the same contour in Dotsenko-Fateev representation.
  For the four-point conformal block we introduce $N_1$ and $N_2$.
  The overall condition (\ref{massrelation4}) reduce these two degree of freedom to one
  which corresponds to the Coulomb modulus of $SU(2)$ theory.
  
  The parameters $m_i$ are the masses associated with the $SU(2)^4 (\subset SO(8))$ flavor symmetry.
  These are related to the masses of the hypermultiplets as
    \bea
    m_1
    &=&    \frac{1}{2} (\mu_1 + \mu_2), ~~~
    m_2
     =     \frac{1}{2} (\mu_3 + \mu_4),
           \nonumber \\
    m_\infty
    &=&    \frac{1}{2} (\mu_1 - \mu_2), ~~~
    m_0
     =     \frac{1}{2} (\mu_3 - \mu_4).
           \label{massSU(2)}
    \eea
  
  The matrix models associated with gauge theories with $N_f=2, 3$ are obtained 
  by taking the decoupling limit of heavy flavors \cite{EM}.
  By taking a limit of $\mu_4 \rightarrow \infty$ while keeping $\mu_4 q = \Lambda_3$ fixed, 
  the matrix model action becomes
    \bea
    W(z)
     =     \mu_3 \log z + m_1 \log (z - 1) - \frac{\Lambda_{3}}{2 z}.
           \label{action3}
    \eea
  with the following condition:
    \bea
    m_1 + m_\infty + \mu_3 + 2 g_s N
     =     0.
           \label{massrelation3}
    \eea
  The prefactor in front of the matrix integral (\ref{matrix}) reduces  to
  $e^{- \frac{m_1 \Lambda_3}{4g_s^2} }$ in this limit.
  This is identified with the (inverse of the) $U(1)$ factor of $N_f=3$ theory (see appendix \ref{subsec:Nf=2inst}).
  
  In order to obtain the $N_f=2$ matrix model, we further take the limit $\mu_2 \rightarrow \infty$ while keeping $\mu_2 \Lambda_3 = \Lambda_2^2$ fixed. The dynamical scale of this gauge theory is
   given by $\Lambda_2$. 
  After rescaling $z \rightarrow \frac{\Lambda_3}{\Lambda_2} z$, 
  the action (\ref{action3}) becomes
    \bea
    W(z)
     =     \mu_3 \log z - \frac{\Lambda_{2}}{2} \left( z + \frac{1}{z} \right).
           \label{action2}
    \eea
  The mass relation reduces in this case to $ \mu_1 + \mu_3 + 2 g_s N = 0$. The prefactor becomes 
  simply $e^{- \frac{\Lambda_2^2}{8g_s^2}}$.

%%%%%%%%%%%%%%%%%%%%%%%%%%%%%%%%%%%%%%%%%%%%%%%%%%%%%%%%%%%%%%%%%%%%%%%%%%%%%%%%%%%%%%%%%%%%%%%%%%%%  section
\section{Planar free energy and prepotential}
  In this section, we will evaluate the planar free energy of the matrix models introduced above.
  In \cite{EM}, we have shown that the free energy of these models satisfies the 
  identities known in Seiberg-Witten thery \cite{Matone, STY, EY}.
  Here, we will evaluate the free energies explicitly and compare them  with the instanton expansions
  of the prepotentials at lower orders.
  The computation is a bit simpler than in the Seiberg-Witten theory
  where both the $A$ and $B$ cycle integrals have to be computed \cite{KLT, IY}. Here  
  we only have to compute the $A$ integral.
  
  We first consider the matrix model for $SU(2)$ gauge theory with $N_f=2$ in next subsection.
  Then, we will analyze the cases of $N_f = 3$ and $4$ theories in turn.
 
%%%%%%%%%%%%%%%%%%%%%%%%%%%%%%%%%%%%%%%%%%%%%%%%%%
\subsection{$SU(2)$ gauge theory with $N_f = 2$}
\label{subsec:Nf=2}
  The matrix model action corresponding to the $SU(2)$ gauge theory with $N_f = 2$ is given by (\ref{action2}).
  For simplicity, we will omit the subscript $2$ of the dynamical scale $\Lambda_2$ below.
  There are two saddle points determined by the classical equation of motion:
    \bea
    W'(z)
     =     \frac{\mu_3}{z} - \frac{\Lambda}{2} \left( 1 - \frac{1}{z^2}  \right)
     =     0.
    \eea
  These lead to the two-cut spectral curve.
  
  The planar loop equation reads as usual 
    \bea
    R(z)
     =   - \frac{1}{2} \left( W'(z) - \sqrt{W'(z)^2 + f(z)} \right),
           \label{resolvent}
    \eea
  where the resolvent is defined by 
  \begin{equation}
  R(z) = \langle \sum_I \frac{g_s}{z - \lambda_I} \rangle.
  \end{equation}
   The function $f$ is given by 
 \begin{equation}
  f(z) = 4g_s \langle \sum_I \frac{W'(z) - W'(\lambda_I)}{z - \lambda_I} \rangle
  = \frac{c_1}{z} + \frac{c_2}{z^2}.
  \end{equation}
 Coefficients $c_1$ and $c_2$ are defined as  
    \bea
    c_1
     =   - 4 g_s \left< \sum_I \left( \frac{\mu_3}{\lambda_I} 
         + \frac{\Lambda}{2 \lambda_I^2} \right) \right>
     =   - 2 g_s N \Lambda,
           ~~~
    c_2
     =   - 2 g_s \left< \sum_I \frac{\Lambda}{\lambda_I} \right>.
    \eea
  In the formula for $c_1$ we have used the equations of motion $\langle \sum_I W'(\lambda_I) \rangle = 0$.
  
  Then, the spectral curve 
  $x^2 = (2 \langle R(z) \rangle + W'(z))^2 = W'(z)^2 + f(z)$
is given by 
    \bea
    x^2
     =     \frac{\Lambda^2}{4 z^4} + \frac{\mu_3 \Lambda}{z^3}
         + \frac{1}{z^2} \left( \mu_3^2 + c_2 - \frac{\Lambda^2}{2} \right)
         + \frac{\mu_1 \Lambda}{z} + \frac{\Lambda^2}{4}.
           \label{spectralNf2}
    \eea
  This is similar to the curve obtained in \cite{GMN}.
  The differential one form is identified with $\lambda_{m} = x dz$
  which has double poles at $t=0$ and $\infty$ with residues  $\mu_3$ and $\mu_1$.
  Note that the parameter $c_2$ corresponds to the variable $u$ in Seiberg-Witten theory.
  
  We evaluate the filling fraction as 
      \bea
    g_s N_1
         =     \frac{1}{4 \pi i} \oint_{C_1} \lambda_{m}(c_2),
           \label{fillingfractionNf2}
    \eea
  where $C_1$ is a cycle around one of the cuts in the curve.
  This integral  is identified with the Coulomb moduli $a$ in the gauge theory 
  and we invert the above relation to solve the unknown parameter $c_2$.
  
    Let us compute the free energy of our model defined by
    \bea
    e^{F_m/g_s^2}
     =     \left( \prod_{I=1}^N \int d \lambda_I \right) \prod_{I < J} (\lambda_I - \lambda_J)^{2} 
           e^{\frac{1}{g_s} \sum_I W(\lambda_I)}.
    \eea
  The starting point is the formula for the $\Lambda$ derivative:
    \bea
    \frac{\partial F_m}{\partial \Lambda}
     =   - \frac{g_s}{2} \left< \sum_{I} \left( \frac{1}{\lambda_I} + \lambda_I \right) \right>
     =     \frac{c_2}{4 \Lambda} - \frac{g_s}{2} \langle \sum_I \lambda_I \rangle.
    \eea
  The expectation value $\langle \sum_I \lambda_I \rangle = \langle \tr M \rangle$ in the second term
  can be determined by studying the large $z$ behavior of the resolvent: 
  $R(z) = - \frac{1}{2} (W'(z) - x) \approx \frac{g_s N}{z} + \frac{g_s \langle \tr M \rangle}{z^2} + \ldots$
    \bea
    g_s \langle \tr M \rangle
     =   - \frac{1}{2 \Lambda} (c_2 - \mu_1^2 + \mu_3^2).
    \eea
  Therefore, we obtain
    \bea
    \Lambda \frac{\partial F_m}{\partial \Lambda}
     =     \frac{1}{4} (2 c_2 - \mu_1^2 + \mu_3^2).
           \label{freeNf2}
    \eea
  Our remaining task is to determine $c_2$ in terms of $g_s N_1$ by using (\ref{fillingfractionNf2}),
  and this leads to the explicit form of the free energy.
  
  To derive $c_2$, let us consider the derivative of (\ref{fillingfractionNf2}) with respect to $c_2$:
    \bea
    4 \pi i \frac{\partial (g_s N_1)}{\partial c_2}
     =     \oint_{C_1} \frac{1}{\Lambda} \frac{dz}{\sqrt{P_4(z)}},
           \label{derivativefillingNf2}
    \eea
  where $P_4$ is the polynomial of degree 4:
    \bea
    P_4(z)
     =     z^4 + \frac{4\mu_1}{\Lambda} z^3 + \frac{4}{\Lambda^2} ( \mu_3^2 + c_2 - \frac{\Lambda^2}{2}) z^2
         + \frac{4\mu_3}{\Lambda} z + 1.
    \eea
  It is easy to transform this polynomial so that 
  (\ref{derivativefillingNf2}) becomes the standard elliptic integral of the first kind.
  In the following, we set $A = \mu_3^2 + c_2 - \frac{\Lambda^2}{2}$ 
  and express the result in terms of $A$. 
  
  For simplicity, let us consider the equal mass case: $\mu_1 = \mu_3 = m$ in the following.
  In this case, by the transformation $z = \frac{t-1}{t+1}$ and rescaling of $t$, 
  the integrand of the right hand side of (\ref{derivativefillingNf2}) can be brought to the standard form 
    \bea
    \frac{\sqrt{2}}{\sqrt{S_+ (\Lambda^2 + 4m \Lambda + 2A)}} \frac{dt}{\sqrt{(1 - t^2)(1 - k^2 t^2)}},
    \eea
  where $k^2 = S_-/S_+$ and 
    \bea
    S_\pm
     =     \frac{1}{\Lambda^2 + 4m \Lambda + 2A} 
           \left( - 3 \Lambda^2 + 2A \pm \Lambda \sqrt{8 \Lambda^2 - 16A + 16m^2} \right).
    \eea
  Then, we can identify the integral (\ref{derivativefillingNf2})
  in terms of the hypergeometric function:
    \bea
    4 \pi i \frac{\partial (g_s N_1)}{\partial A}
    &=&    \frac{2 \sqrt{2}}{\sqrt{S_+ (\Lambda^2 + 4m \Lambda + 2A)}} 
           \int_{1/k}^1 \frac{dt}{\sqrt{(1 - t^2)(1 - k^2 t^2)}}
           \nonumber \\
    &=&    \frac{\sqrt{2} \pi i}{\sqrt{S_+ (\Lambda^2 + 4m \Lambda + 2A)}} 
           F(\frac{1}{2}, \frac{1}{2}, 1; 1 - k^2).
    \eea
  where we have used $\int_{1/k}^1 \frac{dt}{\sqrt{(1 - t^2)(1 - k^2 t^2)}}= iK'(k) = iK (k')$ with $k'^2 = 1 - k^2$.
  We express the right hand side as a small $\Lambda$ expansion which corresponds to the  instanton expansion in gauge theory.
  (Note that $k^2 = 1 + \CO(\Lambda)$.)
  After integrating over $A$, we obtain
    \bea
    2 g_s N_1
    &=&    \sqrt{A}
           \Bigg( 1 - \frac{m^2}{4A^2} \Lambda^2 - \frac{A^2 - 6 A m^2 + 15 m^4}{64 A^4} \Lambda^4 
         - \frac{5 m^2 (3 A^2 - 14 A m^2 + 21 m^4)}{256 A^6} \Lambda^6 
           \nonumber \\
    & &  - \frac{15(A^4 - 28 A^3 m^2 + 294 A^2 m^4 - 924 A m^6 + 1001 m^8)}{16384 A^8} \Lambda^8 
         + \CO(\Lambda^{10}) \Bigg).
    \eea
  Then, we invert this equation and solve for $A$:
    \bea
    A
    &=&     a^2 + \frac{m^2}{2a^2} \Lambda^2
         + \frac{a^4 - 6 m^2 a^2 + 5 m^4}{32 a^6} \Lambda^4
         + \frac{m^2 (5 a^4 - 14 m^2 a^2 + 9m^4)}{64 a^{10}} \Lambda^6
           \nonumber \\
    & &  + \frac{5 a^8 - 252 m^2 a^6 + 1638m^4 a^4 - 2860 m^6 a^2 + 1469 m^8}{8192 a^{14}} \Lambda^8
         + \CO(\Lambda^{10}),
    \eea
  where we have introduced $a = 2g_s N_1$.
  Finally, we substitute this into (\ref{freeNf2}) and integrate by $\Lambda$ to obtain
    \bea
    4F_m
    &=&    2 \left( a^2 - m^2 \right) \log \Lambda
         + \frac{a^2 + m^2}{2 a^2} \Lambda^2 
         + \frac{a^4 - 6 m^2 a^2 + 5 m^4 }{64 a^6} \Lambda^4
         + \frac{m^2 (5 a^4 - 14 m^2 a^2 + 9 m^4)}{192 a^{10}} \Lambda^6
           \nonumber \\
    & &    ~~~~
         + \frac{5 a^8 - 252m^2 a^6 + 1638m^4 a^4 - 2860 m^6 a^2 + 1469 m^8}{32768 a^{14}} \Lambda^8
         + \CO(\Lambda^{10}).
    \eea
  This agrees with the $U(2)$ gauge theory prepotential with $\vec{a} = (a, -a)$ 
  obtained from the Nekrasov partition function (\ref{prepotentialNf=2}) 
  or from the Seiberg-Witten theory \cite{IY}.
  (The first term is the one-loop part and the others are the instanton part.)
  Together with the prefactor $e^{- \frac{\Lambda_2^2}{8g_s^2}}$  
 we see that the full free energy is the same as that of  $SU(2)$ gauge theory.

%%%%%%%%%%%%%%%%%%%%%%%%%%%%%%%%%%%%%%%%%%%%%%%%%%%%%%%%%%%%%%%%%%%%%%%%%%%%%%%%%%%%%%%%%%%%%%%%%%%%  section 2
\subsection{$SU(2)$ gauge theory with $N_f = 3$}
\label{subsec:Nf=3}
  Next, let us consider the matrix model corresponding to the gauge theory with $N_f = 3$.
  The matrix model action is given by (\ref{action3}).
  We will omit the subscript $3$ of the dynamical scale $\Lambda_3$ from now on.
  As in the previous subsection, there are two saddle points in the classical equation of motion.
  In the planar limit, the loop equation leads to the spectral curve
  $x(z)^2 = W'(z)^2 + f(z)$ 
  where $f(z)$ is written as
    \bea
    f(z)
     =     \frac{c_1}{z} + \frac{c_2}{z - 1} + \frac{c_3}{z^2},
    \eea
  with coefficients  
    \bea
    c_1
     =   - 4 g_s \left< \sum_I \left( \frac{\mu_3}{\lambda_I} 
         + \frac{\Lambda}{2 \lambda_I^2} \right) \right>, 
           ~~
    c_2
     =   - 4 g_s \left< \sum_I \frac{m_1}{\lambda_I - 1} \right>,
           ~~
    c_3
     =   - 2 g_s \left< \sum_I \frac{\Lambda}{\lambda_I} \right>.
           \label{f123}
    \eea
  We can easily see that $c_1 + c_2 = 0$ due to the equations of motion.
  
  The one form defined by $\lambda_m \equiv x(z) dz$ 
  has a double pole at $z = 0$ and a simple pole at $z = 1$ and $\infty$
  with residues $\mu_3$, $m_1$ and $m_\infty$, respectively.
  The residue at $z = \infty$ gives a further constraint on $c_i$:
    \bea
    c_2 + c_3
     =     m_\infty^2 - (\mu_3 + m_1)^2.
    \eea
  This condition together with the relation $c_1 + c_2 = 0$ leaves only one of the parameters independent. Let us choose $c_3$ to be independent.  
  
  It is then related to the filling fraction by the integral
    \bea
    4 \pi i g_s N_1
     =     \oint_{C_1} \lambda_{m}(c_3).
           \label{fillingfractionNf3}
    \eea
  For completeness, let us write down here the explicit form of the curve $x^2 = \frac{P_4(z)}{4z^4(z-1)^2}$ with
    \bea
    P_4(z)
    &=&    4 m_\infty^2 z^4 + 4 ((\mu_3 + m_1)\Lambda + m_1^2 - \mu_3^2 - m_\infty^2 - c_3)z^3
           \nonumber \\
    & &  + (\Lambda^2 - 8 \Lambda \mu_3 + 4 \mu_3^2 - 4 \Lambda m_1 + 4 c_3) z^2
         - 2 \Lambda (\Lambda - 2 \mu_3)z + \Lambda^2.
    \eea
  It is convenient to introduce  the notation $B$ as
    \bea
    B= c_3 - \mu_3 \Lambda + \mu_3^2.
    \eea
  The polynomial is then rewritten as 
    \bea
    P_4(z)
    &=&    4 m_\infty^2 z^4 + 4 (\Lambda m_1 + m_1^2 - m_\infty^2 - B)z^3 
         + (\Lambda^2 - 4 \Lambda (\mu_3 + m_1) + 4B) z^2
           \nonumber \\
    & &  - 2 \Lambda (\Lambda - 2 \mu_3)z + \Lambda^2.
    \eea
  
  Let us consider the free energy of this matrix model.
  From the definition, its derivative in $\Lambda$ is written as
    \bea
    \frac{\partial F_m}{\partial \Lambda}
     =   - \frac{g_s}{2} \left< \sum_I \frac{1}{\lambda_I} \right>
     =     \frac{c_3}{4 \Lambda}
     =     \frac{1}{4 \Lambda} (B + \mu_3 \Lambda - \mu_3^2).
           \label{freeenergyNf3}
    \eea
  In order to determine $B$ we take a derivative of (\ref{fillingfractionNf3}) with respect to $B$:
    \bea
    4 \pi i \frac{\partial (g_s N_1)}{\partial B}
     =   - \oint_{C_1} \frac{dz}{\sqrt{P_4(z)}}.
           \label{derivativeNf3}
    \eea
  For simplicity, we consider the case where $\mu_3 = m$ and $m_1 = m_\infty = 0$ in what follows.
  In this case, $P_4$ becomes a polynomial of degree 3:
    \bea
    P_3(z)
     =     (z-1) (- 4 Bz^2 + (\Lambda^2 - 4 \Lambda m)z  - \Lambda^2).
    \eea
  After a change of variable (first shifting $z \rightarrow z - p$ and then rescaling as $z = Q t$), we obtain
    \bea
    P_3(z)
     \rightarrow
         - 4 B Q^2 (1 + p) \times t (1 - t) (1 - k^2 t),
    \eea
  where
    \bea
    k^2
     =     \frac{Q}{1 + p}, ~~
    p
     =     \frac{1}{2} \left( - \frac{\Lambda}{4B} (\Lambda - 4m) + Q \right), ~~
    Q
     =     \frac{\Lambda}{4B} \sqrt{(\Lambda - 4m)^2 - 16 B}.
    \eea
  
  As a result, (\ref{derivativeNf3}) becomes
    \bea
    4 \pi i \frac{\partial (g_s N_1)}{\partial B}
    &=&  - \frac{1}{\sqrt{-B(1 + p)}} \int_0^1 \frac{dt}{\sqrt{t (1 - t) (1 - k^2 t)}}
           \nonumber \\
    &=&  - \frac{\pi}{\sqrt{-B(1 + p)}} F(\frac{1}{2}, \frac{1}{2}, 1; k^2).
    \eea
  By expanding the hypergeometric function and then integrating over $B$, we obtain
    \bea
    2 g_s N_1
    &=&    \sqrt{B} \Bigg( 1 + \frac{m \Lambda}{4 B} - \frac{1}{64B^2} (B + 3m^2) \Lambda^2
         + \frac{m}{256B^3} (5m^2 + B) \Lambda^3 
           \nonumber \\
    & &    - \frac{1}{16384 B^4}(3B^2 + 30m^2 B + 175 m^4) \Lambda^4
         - \frac{m}{65536B^5}(9B^2 + 70m^2 B + 441m^4) \Lambda^5
           \nonumber \\
    & &  - \frac{1}{1048576 B^6} (5 B^3 + 105m^2 B^2 + 735 m^4B + 4851 m^6) \Lambda^6 + \CO(\Lambda^7) \Bigg).
    \eea
  We invert this equation for $B$, 
    \bea
    B
    &=&     a^2 - \frac{m \Lambda}{2} + \frac{m^2 + a^2}{32a^2} \Lambda^2
          + \frac{a^4 - 6a^2 m^2 + 5 m^4}{8192a^6} \Lambda^4
          + \frac{m}{16384a^8} (9a^4 + 70m^2a^2 +441m^4) \Lambda^5
            \nonumber \\
    & &   + \frac{m^2}{262144 a^{10}} (185a^4 + 1946m^2 a^2 + 15885 m^4) \Lambda^6 + \CO(\Lambda^7), 
    \eea
  where we have defined $a = 2 g_s N_1$.
  Finally, by substituting this into (\ref{freeenergyNf3}), we obtain
    \bea
    4F_m
    &=&    (a^2 - m^2) \log \Lambda + \frac{m \Lambda}{2} + \frac{m^2 + a^2}{64 a^2} \Lambda^2
         + \frac{a^4 - 6 m^2 a^2 +5 m^4}{2^{15} a^6} \Lambda^4
           \nonumber \\
    & &  + \frac{m}{2^{14} \times 5} \frac{9a^4 + 70 m^2 a^2 +441 m^4}{a^8} \Lambda^5
         + \frac{m^2}{2^{19} \times 3} \frac{185a^4 + 1946 m^2 a^2 + 15885 m^4}{a^{10}} \Lambda^6 + \CO(\Lambda^7).
           \nonumber \\
    & &    ~~~~~~~~
    \eea
  Term with $\log \Lambda$ is the one-loop contribution.
 Remaining terms agree precisely with the prepotential obtained from the Nekrasov partition function (\ref{prepotentialNf=3}).

%%%%%%%%%%%%%%%%%%%%%%%%%%%%%%%%%%%%%%%%%%%%%%%%%%
\subsection{$SU(2)$ gauge theory with $N_f=4$}
\label{subsec:Nf4}
  We now consider the matrix model with the original action (\ref{action4}).
  The planar loop equation 
$ R(z)=-{1\over 2}\left(W'(z)-\sqrt{W'(z)^2+f(z)}\right)$ 
involves a function $f(z)$ which now has a form
  $f(z) = \sum_{i=0}^2 \frac{c_i}{z-q_i}$. Parameters $\{c_i\}$ are given by 
  \begin{equation}
  c_0=-4g_sm_0\langle \,\sum_I {1\over \lambda_I}\,\rangle,\hskip2mm 
   c_1=-4g_sm_1\langle \,\sum_I {1\over \lambda_I-1}\,\rangle,\hskip2mm 
 c_2=-4g_sm_2\langle \,\sum_I {1\over \lambda_I-q}\,\rangle.
  \end{equation}
 By studying the behavior of loop equation at large $z$ we find that the parameters obey 
    \bea
    \sum_{i=0}^2 c_i
     =     0, ~~
    c_1 + qc_2
     =     m_\infty^2 - (\sum_{i=0}^2 m_i)^2.
           \label{c}
    \eea
  By eliminating $c_1$ and $c_2$, the spectral curve becomes
    \bea
    x^2
     =     \frac{P_4(z)}{z^2(z-1)^2(z-q)^2},
    \eea
  where $P_4$ is the following polynomial of degree $4$
    \bea
    P_4(z)
    &=&    m_\infty^2 z^4 
         + \Big( -(1+q)(m_\infty^2+ m_0^2) + (1-q) (m_1^2 - m_2^2) - 2 m_0 (q m_1 + m_2) + q c_0 \Big) z^3
           \nonumber \\
    & &  + \Big( q m_\infty^2 + (1 + 3q + q^2)m_0^2 + (1-q)(m_2^2-qm_1^2) + 2(1+q)m_0 (qm_1 + m_2)
           \nonumber \\
    & &  - (1+q)q c_0 \Big) z^2
         + \Big( -2q(1+q)m_0^2 - 2q^2m_0m_1 - 2qm_0m_2 + q^2 c_0 \Big)z + q^2 m_0^2.
    \eea
  The meromorphic one form $xdz$ has simple poles at $z = 0, 1, q$ and $z=\infty$ 
  with residues $m_0,m_1,m_2$ and $m_\infty$.
  
  Again, we consider the derivative of the free energy:
    \bea
    \frac{\partial F_m}{\partial q}
     =     g_s m_2 \left< \tr \frac{1}{q - M} \right>
     =     m_2  R(z) |_{z=q}.
    \eea
  This can be easily computed by expanding the resolvent at $z = q$,
  $ R(z)  = \frac{c_2}{4m_2} + \CO(z-q)$.
  Then, we obtain a simple expression for the free energy
    \bea
    \frac{\partial F_m}{\partial q}
     =     \frac{c_2}{4}
     =     \frac{1}{4(1-q)} \left( (\sum_{i=0}^2 m_i)^2 - m_\infty^2 - c_0 \right).
           \label{freeenergyderivative4}
    \eea
  In the last equality we  used the relation (\ref{c}).
  
  In what follows, we consider the simple case where all the hypermultiplet masses are equal to $m$:
  i.e. $m_0 = m_\infty = 0$ and $m_1 = m_2 = m$.
  In this case, the polynomial is reduced to degree $3$: $P_3(z) = C z (z - z_+)(z - z_-)$, 
  where we have introduced $C \equiv c_0 q$ and 
    \bea
    z_\pm
     =     \frac{1}{2} \left( 1+q - (1-q)^2 \frac{m^2}{C} 
         \pm (1-q) \sqrt{1 - 2(1+q) \frac{m^2}{C} + (1-q)^2 \frac{m^4}{C^2} } \right).
    \eea
  By taking the $C$ derivative of $xdz$,  the holomorphic one form 
  becomes 
    \bea
    \frac{\partial}{\partial C} xdz
     =     \frac{1}{2 \sqrt{C z_+}} \frac{dz}{\sqrt{z (1-z)(1 - k^2 z)}}, ~~~
    k^2
     =     \frac{z_-^2}{q}.
    \eea
  The remaining calculation is similar to those considered in the previous subsections.
  That is, we first evaluate the period integral of the above one form.
  Then by expanding in $\frac{m^2}{C}$ and integrating over $C$, we obtain
    \bea
    2 i g_s N_1
     =     \sqrt{C} \left( h_0(q) - h_1(q) \frac{m^2}{C} - \frac{h_2(q)}{3} \frac{m^4}{C^2} 
         - \frac{h_3(q)}{5} \frac{m^6}{C^3} + \CO(\frac{m^8}{C^4}) \right),
    \eea
  where $h_i(q)$ are the expansion coefficients of the period integral 
  in $\frac{m^2}{C}$ and depend only on $q$.
  $h_0(q)$  is for the theory with massless flavors:
    \bea
    h_0(q)
     =     1 + \frac{1}{4} q + \frac{9}{64} q^2 + \frac{25}{256} q^3 + \frac{1225}{16384}q^4 + \CO(q^5).
           \label{B_0}
    \eea
  Lower order expansions of $h_1, h_2$ and $h_3$ are given by 
    \bea
    h_1(q)
    &=&    \frac{1}{2} + \frac{1}{8} q + \frac{1}{128} q^2 + \frac{1}{512} q^3 + \frac{25}{32768} q^4 + \CO(q^5),
           \nonumber \\
    h_2(q)
    &=&    \frac{3}{8} + \frac{27}{32} q + \frac{27}{512} q^2 + \frac{3}{2048} q^3 + \frac{27}{131072} q^4 + \CO(q^5),
           \nonumber \\
    h_3(q)
    &=&    \frac{5}{16} + \frac{125}{64} q + \frac{1125}{1024} q^2 + \frac{125}{4096} q^3 + \frac{125}{262144} q^4
         + \CO(q^5).
    \eea
  Solving for $C$, we obtain
    \bea
    C
    &=&    a^2 \Bigg( \frac{1}{h_0(q)^2} + \frac{2h_1(q)}{h_0(q)} \frac{m^2}{a^2}
         + \frac{2 h_0(q) h_2(q) - 3 h_1(q)^2}{3} \frac{m^4}{a^4} 
           \nonumber \\
    & &    ~~~~~
         + \frac{10 h_0(q) h_1(q)^3 - 10 h_0(q)^2 h_1(q) h_2(q) + 2 h_0(q)^3 h_3(q)}{5} \frac{m^6}{a^6}
         + \ldots \Bigg),
    \eea
  where $a = 2 i g_s N_1$.
  By substituting the above expression into (\ref{freeenergyderivative4}) and integrating over $q$, we finally obtain the $N_f=4$ free energy
    \bea
    4F_m
    &=&    (a^2 - m^2) \log q + \frac{a^4+6 a^2 m^2 + m^4}{2 a^2} q
         + \frac{13 a^8 + 100 m^2 a^6 + 22 m^4 a^4 - 12 m^6 a^2 + 5 m^8}{64a^6}q^2
           \nonumber \\
    & &  + \frac{23 a^{12} + 204 m^2 a^{10} + 51 m^4 a^8 - 48 m^6 a^6 + 45 m^8 a^4 - 28 m^{10} a^2 + 9 m^{12}}{192a^{10}} q^3
         + \CO(q^4).
    \eea
  This perfectly agrees with the instanton partition function (\ref{prepotentialNf=4}).
  
  Finally, we make a brief comment on the massless theory.
  In this case, the expression for $C$ simplifies and becomes $C=a^2/h_0^2(q)$ where $h_0(q)$ is (\ref{B_0}).
  Thus, it is easy to derive 
    \bea
    4 F_m
     =     a^2 \left( \log q - \log 16 + \frac{1}{2} q + \frac{13}{64} q^2 + \frac{23}{192} q^3 + \frac{2701}{32768} q^4
         + \frac{5057}{81920} q^5 + \CO(q^6) \right),
    \eea
  where we have added the one-loop contribution $- a^2 \log 16$.
  Note that this can be written as $4 F_m = a^2 \log q_{{\rm IR}}$ 
  where $q$ and $q_{{\rm IR}}=e^{2\pi i\tau_{{\rm IR}}}$ are related by
    \bea
    q 
     =     \frac{\vartheta_2(\tau_{{\rm IR}})^4}{\vartheta_3(\tau_{{\rm IR}})^4}
     =     16 q_{{\rm IR}} - 128 q_{{\rm IR}}^2 + 704 q_{{\rm IR}}^3 - 3072 q_{{\rm IR}}^4 +11488 q_{{\rm IR}}^5
         + \ldots.
    \eea
  as already discussed in \cite{DKM, GKMW, AGT, MMMcoupling, Poghossian, EM}.
  Thus the theory appears classical in terms of IR coupling constant $\tau_{{\rm IR}}$.
  
%%%%%%%%%%%%%%%%%%%%%%%%%%%%%%%%%%%%%%%%%%%%%%%%%%%%%%%%%%%%%%%%%%%%%%%%%%%%%%%%%%%%%%%%%%%%%%%%%%%%  section
\section{Matrix model and Quiver gauge theories}
\label{sec:quiver}
  In this section, we study matrix models which describe $\CN=2$ $SU(2)$ quiver gauge theories.
  First of all, we consider a matrix model describing $SU(2)$ linear quiver gauge theory 
  where each gauge group has a vanishing beta function \cite{Gaiotto}.
  Then by taking its decoupling limit, we propose models for asymptotically free gauge theories in subsection \ref{subsec:asymptotically}.
  
\begin{figure}[t]
    \begin{center}
    \includegraphics[scale=0.7]{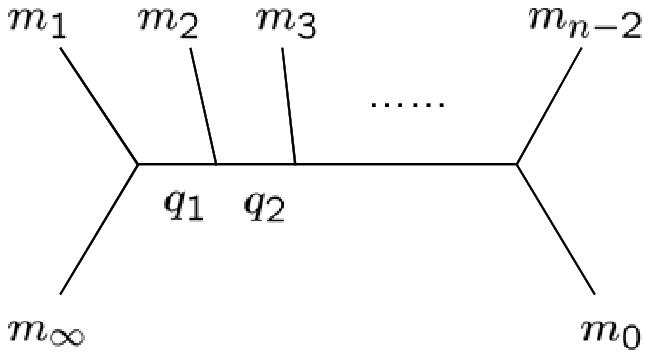}
    \caption{{\small }}
    \label{fig:quivern-2}
    \end{center}
    \end{figure}

According to the AGT conjecture, $SU(2)^{n-3}$ linear quiver gauge theory is related to the $n$-point conformal block of the Liouville theory, 
  which is represented by the trivalent graph \cite{BPZ} as in Fig \ref{fig:quivern-2}.
  As seen in section \ref{sec:matrix}, the Dotsenko-Fateev representation of the conformal block suggests a 
  matrix model 
  with the following action \cite{DV}:
    \bea
    W(M) 
     =     \sum_{i = 0}^{n-2} m_i \log (M - t_i), 
    \eea
  where $t_0 = 0$ and $t_1 = 1$.
  Other parameters $t_i = \prod_{k=1}^{i-1} q_k$ ($i = 2, \ldots, n-2$) describe complex structure 
  of the $n$-punctured sphere.
  Note that we also have the prefactor as in (\ref{matrix})
    
  From the gauge theory perspective, the parameters $q_k$ are related 
  to the gauge coupling constants $q_k = e^{2 \pi i \tau_k}$ of the gauge group $SU(2)^{n-3}$.
  For $n=4$, this reduces to the matrix model which we studied in subsection \ref{subsec:Nf4}.
  Parameters $m_0$ and $m_{n-2}$ are related to the mass parameters of two hypermultiplets
  in the fundamental representation of the $SU(2)$ at one end of the quiver: $m_{n-2}=\frac{1}{2}(\mu_3 + \mu_4)$
  and $m_0 = \frac{1}{2} (\mu_3 - \mu_4)$.
  Also, the masses of the bifundamental hypermultiplets are identified with $m_i$ ($i=2, \ldots, n-3$).
  Finally, the masses of two fundamental hypermultiplets of the other end of the quiver are related
  to $m_1$ and $m_\infty$ as $m_1=\frac{1}{2}(\mu_1 + \mu_2)$ and $m_\infty=\frac{1}{2}(\mu_1 - \mu_2)$.
  The mass parameter $m_\infty$ is introduced by the following condition:
    \bea
    \sum_{i=0}^{n-2} m_i + m_\infty + 2 g_s N 
     =     0.
           \label{massrelationquiver}
    \eea
  The critical points are determined by the equation of motion 
    \bea
    \sum_{i = 0}^{n-2} \frac{m_i}{\lambda_I - t_i}
         + 2 g_s \sum_{J (\neq I)} \frac{1}{\lambda_I - \lambda_J}
     =     0.
           \label{eom}
    \eea
  If we ignore the second term, we obtain $n-2$ critical points $e_p$ ($p = 1, 2, \ldots, {n-2}$).
  Let each $N_p$ ($p = 1, 2, \ldots, {n-2}$) be the number of the matrix eigenvalues which are at the critical point $e_p$.
  
  We take the large $N$ limit with mass parameters $\{m_i\}$ and filling fractions $\{\nu_p \equiv g_s N_p\}$ being kept fixed.
  Since this is one matrix model, the loop equation is still the same as in the previous cases (\ref{resolvent})
    \bea
    f(z) 
     \equiv
           4 g_s \tr \left< \frac{W'(z) - W'(M)}{z - M}  \right>
     =     \sum_{i = 0}^{n-2} \frac{c_i}{z - t_i}
     \equiv
           \frac{Z(t)}{\prod_{i=0}^{n-2} (z - t_i)}.
           \label{fi}
    \eea
  We note that a polynomial $Z(t)$ is of degree $n-3$, 
  since the leading term vanishes due to equations of motion.
  
  Finally, we define the meromorphic one form $\lambda = x(z) dz$ as
    \bea
    x(z)^2
     \equiv     
           \left( 2 \langle R(z) \rangle + W'(z) \right)^2
     =     W'(z)^2 + f(z).
           \label{spectral}
    \eea

%%%%%%%%%%%%%%%%%%%%%%%%%%%%%%%%%%%%%%%%%%%%%%%%%%
\subsection{Matrix model for asymptotically free quiver gauge theory}
\label{subsec:asymptotically}
  The matrix model corresponding to asymptotically free quiver gauge theory can be obtained 
  by taking the decoupling limit as in section \ref{sec:matrix}.
  Only possible limits which does not spoil the condition (\ref{massrelationquiver}) is
  the case where $\mu_2 (=m_1 - m_\infty)$ or $\mu_4 (=m_{n-2} - m_0)$ is taken to infinity.
  
  For the sake of illustration, let us consider the $n=5$ case with the action
    \bea
    W(z) 
     =     \sum_{i = 0}^{3} m_i \log (z - t_i),
    \eea
  where $t_2 = q_1$ and $t_3 = q_1 q_2$.
  This corresponds to $SU(2)_1 \times SU(2)_2$ quiver gauge theory whose gauge coupling constants are $q_1$ and $q_2$.
  We first take a limit $\mu_4 \rightarrow \infty$ with $\mu_4 q_2 = \tilde{\Lambda}$ fixed.
  In this limit, we obtain
    \bea
    W(z)
     \rightarrow
           \mu_3 \log z + \sum_{i=1,2} m_i \log (z - t_i) - \frac{q_1 \tilde{\Lambda}}{2z}.
           \label{asymptoticallyquiver1}
    \eea
  It is natural to anticipate that this matrix model corresponds to the quiver theory 
  of one fundamental matter coupled to the second gauge group $SU(2)_2$
  and two fundamental multiplets are coupled to the first gauge group.
  The relation of the mass parameters (\ref{massrelationquiver}) becomes 
  $\mu_3 + \sum_{i=1,2} m_i + m_\infty + 2 g_s N =0$ in this limit.
  
  By further taking the limit $\mu_2 \rightarrow \infty$ with $\mu_2 q_1 = \Lambda$ fixed, 
  we obtain from (\ref{asymptoticallyquiver1})
    \bea
    W(z)
     \rightarrow
           \mu_3 \log z + m_2 \log (z - 1) - \frac{\Lambda z}{2} - \frac{\tilde{\Lambda}}{2z}, 
    \eea
  where we have also rescaled $z \rightarrow q_1 z$. 
  The relation of the mass parameters (\ref{massrelationquiver}) becomes
    \bea
    \mu_3 + m_2 + \mu_1 + 2 g_s N
     =     0.
    \eea
  This matrix model is expected to describe $SU(2)_1 \times SU(2)_2$ quiver gauge theory with
  each gauge factor coupled to one hypermultiplet.
  Both of the gauge factors have nonvanishing beta functions and the theory is asymptotically free.
  
  It is possible to generalize this construction to the case with $n>5$.
  A decoupling limit of a hypermultiplet  
  at the last end of the quiver is $\mu_4 \rightarrow \infty$ with $\mu_4 q_{n-3} = \tilde{\Lambda}$ fixed.
  Also, another decoupling limit of a hypermultiplet at the first end of the quiver
  is $\mu_2 \rightarrow \infty$ with $\mu_2 q_{1} = \Lambda$ fixed.
  By taking these limits, we finally obtain
    \bea
    W(z)
     =     \mu_3 \log z + m_2 \log (z - 1) + \sum_{i=3}^{n-3} m_i \log \left(z - \prod_{k=2}^{i-1} q_k \right)
         - \frac{\Lambda z}{2} - \frac{\tilde{\Lambda}}{2z} \left( \prod_{k=2}^{n-4}q_k \right),
    \eea
  with the following relation for the mass parameters:
    \bea
    \mu_3 + \sum_{i=2}^{n-3} m_i + \mu_1 + 2 g_s N
     =     0.
    \eea

%%%%%%%%%%%%%%%%%%%%%%%%%%%%%%%%%%%%%%%%%%%%%%%%%%%%%%%%%%%%%%%%%%%%%%%%%%%%%%%%%%%%%%%%%%%%%%%%%%% section 
\section{Conclusion and discussion}
 
 In this paper we have studied the matrix model proposed to explain the AGT relation and 
 interpolate the Liouville and $\CN=2$ $SU(2)$ gauge theories. We have explicitly evaluated the free energy of the matrix models describing $SU(2)$ gauge theory with $N_f=2,3,4$ flavors and have shown that they in fact reproduce the 
amplitudes of Seiberg-Witten theory. Our analysis is limited to the large $N$ limit and 
it is very important to see if our results can be generalized and reproduce full Nekrasov partition functions. 
There is already an interesting work in this direction \cite{FHT, KPW} and 
we hope that we can report further results in future publications.

%%%%%%%%%%%%%%%%%%%%%%%%%%%%%%%%%%%%%%%%%%%%%%%%%%%%%%%%%%%%%%%%%%%%%%%%%%%%%%%%%%%%%%%%%%%  acknowledgements
\section*{Acknowledgements}
  K.M. would like to thank K.~Hosomichi, H.~Itoyama and F.~Yagi
  for discussions and comments. 
  He also would like to thank Ecole normale Superieure,  
  SISSA and University of Amsterdam 
  for warm hospitality during part of this project.
  Research of T.E. is supported in part by the project 19GS0219
  of the Japan Ministry of
  Education, Culture, Sports, Science and Technology.
  Research of K.M. is supported in part by JSPS Bilateral Joint Projects (JSPS-RFBR collaboration).

%%%%%%%%%%%%%%%%%%%%%%%%%%%%%%%%%%%%%%%%%%%%%%%%%%%%%%%%%%%%%%%%%%%%%%%%%%%%%%%%%%%%%%%%%%%%%%%%%%%%%
\appendix

\section*{Appendix}
%%%%%%%%%%%%%%%%%%%%%%%%%%%%%%%%%%%%%%%%%%%%%%%%%%%%%%%%%%%%%%%%%%%%%%%%%%%%%%%%%%%%%%%%%%%%%%%%%%% 
\section{Nekrasov partition function}
\label{sec:Nekrasov}
  The instanton partition function of $\CN=2$ $U(2)$ gauge theory with $N_f = 4$ is expressed 
  as a sum over all possible Young tableaus parametrized as $Y = (\lambda_1 \geq \lambda_2 \geq \ldots)$
  where $\lambda_\ell$ is the height of the $\ell$-th column \cite{Nekrasov, AGT}:
    \bea
    Z_{{\rm inst}}
    &=&    \sum_{(Y_1, Y_2)} q^{|\vec{Y}|}
           Z_{{\rm vector}} (\vec{a}, \vec{Y})
           Z_{{\rm antifund}} (\vec{a}, \vec{Y}, \mu_1)
           Z_{{\rm antifund}} (\vec{a}, \vec{Y}, \mu_2)
           Z_{{\rm fund}} (\vec{a}, \vec{Y}, -\mu_3)
           Z_{{\rm fund}} (\vec{a}, \vec{Y}, -\mu_4).
           \nonumber \\
           \label{ZinstU(2)}
    \eea
  Here
    \bea
    Z_{{\rm vector}} (\vec{a}, \vec{Y})
    &=&    \prod_{i,j = 1,2} \prod_{s \in Y_i}
           \left( a_{ij} - \epsilon_1 L_{Y_j}(s) + \epsilon_2 (A_{Y_i}(s) + 1) \right)^{-1}
           \nonumber \\
    & &    ~~~~~~~~~~~~~~~\times
           \prod_{t \in Y_j} \left( a_{ji} + \epsilon_1 L_{Y_j}(t) - \epsilon_2 (A_{Y_i}(t) + 1) + \epsilon_+ \right)^{-1},
           \nonumber \\
    Z_{{\rm fund}} (\vec{a}, \vec{Y}, \mu)
    &=&    \prod_{i = 1,2} \prod_{s \in Y_i}
           \left( a_i + \epsilon_1 (\ell - 1) + \epsilon_2 (m - 1) - \mu + \epsilon_+ \right),
           \nonumber \\
    Z_{{\rm antifund}} (\vec{a}, \vec{Y}, \mu)
    &=&    \prod_{i = 1,2} \prod_{s \in Y_i}
           \left( a_i + \epsilon_1 (\ell - 1) + \epsilon_2 (m - 1) + \mu \right),
    \eea 
  and $\epsilon_+ = \epsilon_1 + \epsilon_2$ and $a_{ij} = a_i - a_j$. 
  For a box $s$ at the coordinate $(\ell, m)$, 
  the leg-length $L_{Y}(s) = \lambda'_m - \ell$ and the arm-length $A_{Y}(s) = \lambda_\ell - m$ 
  where $\lambda'_m$ is the length of the $m$-th row.
  The minus signs of the masses in $Z_{{\rm fund}}$ are due to the convention.
  
  In order to derive the expression for $SU(2)$ gauge theory, we set the Coulomb moduli as $\vec{a} = (a, - a)$ 
  which gives
    \bea
    Z_{{\rm vector}} (a, \vec{Y})
    &=&    \prod_{i = 1,2} \prod_{s \in Y_i}
           \left( 2 a \delta_{ij} - \epsilon_1 L_{Y_j}(s) + \epsilon_2 (A_{Y_i}(s) + 1) \right)^{-1}
           \nonumber \\
    & &    ~~~~~~~~~~~~~~~~~\times
           \prod_{t \in Y_j} \left( - 2 a \delta_{ij} + \epsilon_1 L_{Y_j}(t)
         - \epsilon_2 (A_{Y_i}(t) + 1) + \epsilon_+ \right)^{-1},
           \nonumber \\
    Z_{{\rm fund}} (a, \vec{Y}, \mu)
    &=&    \prod_{i = 1,2} \prod_{s \in Y_i}
           \left( a \delta_i + \epsilon_1 (\ell - 1) + \epsilon_2 (m - 1) - \mu + \epsilon_+ \right),
           \nonumber \\
    Z_{{\rm antifund}} (a, \vec{Y}, \mu)
    &=&    \prod_{i = 1,2} \prod_{s \in Y_i}
           \left( a \delta_i + \epsilon_1 (\ell - 1) + \epsilon_2 (m - 1) + \mu \right),
    \eea 
  where we define $\delta_1 = +1$ and $\delta_2 = -1$, and
    \bea
    \delta_{ij}
     =     \left\{ \begin{array}{ll}
           0 & {\rm for}~ i = j, \\
           1 & {\rm for}~ i = 1 ~{\rm and}~ j = 2, \\
         - 1 & {\rm for}~ i = 2 ~{\rm and}~ j = 1. \\
           \end{array} \right.
    \eea
  Then, the $SU(2)$ and $U(2)$ partition functions are related by the $U(1)$ factor as pointed out in \cite{AGT}:
    \bea
    Z_{{\rm inst}}|_{\vec{a}=(a, -a)}
     =     f^{U(1)} Z_{{\rm inst}}^{SU(2)}, ~~~
    f^{U(1)}
     =     (1-q)^{\frac{\mu_1 + \mu_2}{\epsilon_1 \epsilon_2} (\epsilon_+ + \frac{\mu_3 + \mu_4}{2})}.
    \eea
  Note that this expression differs by  a minus sign in front of $(\mu_3 + \mu_4)$ from the one of \cite{AGT}.
   As argued in \cite{AGT}, the $SU(2)$ partition function is invariant under  ``flips".
  These flips are reduced in the self-dual case $\epsilon_1 = - \epsilon_2 = \hbar$ to 
    \bea
    a 
     \rightarrow
        - a,~~~
    \mu_1 \pm \mu_2 
     \rightarrow
         - (\mu_1 \pm \mu_2),~~~
    \mu_3 \pm \mu_4
     \rightarrow 
         - (\mu_3 \pm \mu_4).
    \eea 
  
  Gauge theory prepotential can be obtained in the limit where the deformation parameters go to zero
  (with a fixed ratio $\epsilon_1/\epsilon_2$):
    \bea
    \CF_{{\rm inst}}
     =     \lim_{\epsilon_{1,2} \rightarrow 0} (- \epsilon_1 \epsilon_2) \log Z_{{\rm inst}}.
    \eea
  In the self-dual case,
  $SU(2)$ gauge theory prepotential is written as
    \bea
    \CF^{SU(2)}_{{\rm inst}}
    &=&    \lim_{\hbar \rightarrow 0} \hbar^2 (\log Z_{{\rm inst}} - \log f^{U(1)})
           \nonumber \\
    &=&    \CF_{{\rm inst}} + \frac{1}{2} (\mu_1 + \mu_2)(\mu_3 + \mu_4) \log (1 - q).
    \eea
  To compare with the free energy of the matrix model, we present an expansion of
  $\CF_{{\rm inst}}$ for the equal mass case $\mu_i=m$
    \bea
    \CF_{{\rm inst}}
    &=&    \frac{a^4 + 6 m^2 a^2 + m^4}{2 a^2} q + \frac{13 a^8 + 100 m^2 a^6 + 22 m^4 a^4 - 12 m^6 a^2 + 5 m^8}{64a^6}q^2
           \label{prepotentialNf=4}\\
    & &  + \frac{23 a^{12} + 204 m^2 a^{10} + 51 m^4 a^8 - 48 m^6 a^6 + 45 m^8 a^4 - 28 m^{10} a^2 + 9 m^{12}}{192a^{10}} q^3
         + \CO(q^4).
           \nonumber
    \eea

%%%%%%%%%%%%%%%%%%%%%%%%%%%%%%%%%%%%%%%%%%%%%%%%%%
\subsection{$U(2)$ gauge theory with $N_f = 3$}
\label{subsec:Nf=3inst}
  Let us consider Nekrasov partition function of the theory with $N_f = 3$.
  This can be obtained from the above partition function by taking a limit $\mu_4 \rightarrow \infty$
  with $\mu_4 q \equiv \Lambda_3$ fixed.
  In the $k$-instanton part the only factor which contains $\mu_4$ is
    \bea
    Z_{{\rm fund}}(a, \vec{Y}, -\mu_4)
     =     \prod_{i = 1,2} \prod_{s \in Y_i}
           \left( a \delta_i + \epsilon_1 (\ell - 1) + \epsilon_2 (m - 1) + \mu_4 + \epsilon_+ \right).
    \eea
  When combined with $k$-instanton factor $q^k$, this gives the leading contribution $\Lambda_3^k$
  and the other contributions are suppressed in the limit.
  Therefore, we obtain
    \bea
    Z_{{\rm inst}}^{N_f=3}
    &=&    \sum_{(Y_1, Y_2)} \Lambda_3^{|\vec{Y}|}
           Z_{{\rm vector}} (a, \vec{Y})
           Z_{{\rm antifund}} (a, \vec{Y}, \mu_1)
           Z_{{\rm antifund}} (a, \vec{Y}, \mu_2)
           Z_{{\rm fund}} (a, \vec{Y}, - \mu_3).
           \nonumber \\
    \eea
  The $U(1)$ factor reduces to
    \bea
    f^{U(1)}
     \rightarrow
    f^{U(1), N_f=3}
     =     \exp \left(- \frac{(\mu_1 + \mu_2) \Lambda_3}{2 \epsilon_1 \epsilon_2} \right).
           \label{U(1)Nf3}
    \eea
  In the simple case of $\mu_3=m$ and 
  $\mu_1=\mu_2=0$ which   we  considered in  subsection \ref{subsec:Nf=3},
  the prepotential of the gauge theory is given by 
    \bea
    \CF^{N_f=3}_{{\rm inst}}
     =     \frac{1}{2} m \Lambda_3 + \frac{a^2 + m^2}{64a^2} \Lambda_3^2 
          + \frac{a^4 - 6 m^2 a^2 +5 m^4}{2^{15} a^6} \Lambda_3^4 + \CO(\Lambda_3^5).
           \label{prepotentialNf=3}
    \eea
    
%%%%%%%%%%%%%%%%%%%%%%%%%%%%%%%%%%%%%%%%%%%%%%%%%%
\subsection{$U(2)$ gauge theory with $N_f = 2$}
\label{subsec:Nf=2inst}
  We can further take a limit where $\mu_2 \rightarrow \infty$ while keeping $\mu_2 \Lambda_3 \equiv \Lambda_2^2$ fixed.
  In this limit, the partition function becomes:
    \bea
    Z_{{\rm inst}}^{N_f=2}
    &=&    \sum_{(Y_1, Y_2)} \Lambda_2^{2|\vec{Y}|}
           Z_{{\rm vector}} (a, \vec{Y})
           Z_{{\rm antifund}} (a, \vec{Y}, \mu_1)
           Z_{{\rm fund}} (a, \vec{Y}, -\mu_3),
    \eea
  and the $U(1)$ factor is reduced to $f^{U(1)} \rightarrow \exp \left(- \frac{\Lambda_2^2}{2 \epsilon_1 \epsilon_2} \right)$.
$SU(2)$ prepotential is given by 
    \bea
    \CF^{SU(2), N_f=2}_{{\rm inst}}
     =     \CF^{N_f=2}_{{\rm inst}} - \frac{\Lambda_2^2}{2}.
    \eea
  For the equal mass case with $\mu_1 = \mu_3 = m$, 
  lower terms of instanton expansion are given by  
    \bea
    \CF^{N_f=2}_{{\rm inst}}
    &=&    \frac{a^2 + m^2}{2a^2} \Lambda_2^2 + \frac{a^4 - 6 a^2 m^2 + 5 m^4}{64a^6} \Lambda_2^4
         + \frac{m^2(5 a^4 - 14 a^2 m^2 + 9 m^4)}{192a^{10}} \Lambda_2^6 + \CO(\Lambda_2^8).
           \nonumber \\
           \label{prepotentialNf=2}
    \eea

%%%%%%%%%%%%%%%%%%%%%%%%%%%%%%%%%%%%%%%%%%%%%%%%%%%%%%%%%%%%%%%%%
%%% references
%%%%%%%%%%%%%%%%%%%%%%%%%%%%%%%%%%%%%%%%%%%%%%%%%%%%%%%%%%%%%%%%%

\end{document}